\documentclass{jetpl}
\twocolumn
\lat
\title{ Measurement of the Pion Form Factor 
in the Range $1.04-1.38$\,GeV with the CMD-2 Detector }
\rtitle{Measurement of the Pion Form Factor 
in the Range $1.04-1.38$\,GeV with the CMD-2 Detector }
\sodtitle{Measurement of the Pion Form Factor 
in the Range $1.04-1.38$\,GeV with the CMD-2 Detector }

\author{
V.\,M.\,Aulchenko$^{a,b}$, R.\,R.\,Akhmetshin$^{a}$,
V.\,Sh.\,Banzarov$^{a}$, L.\,M.\,Barkov$^{a,b}$,
N.\,S.\,Bashtovoy$^{a}$, D.\,V.\,Bondarev$^{a,b}$,
A.\,E.\,Bondar$^{a}$, A.\,V.\,Bragin$^{a}$,
A.\,A.\,Valishev$^{a}$,  N.\,I.\,Gabyshev$^{a}$,
D.\,A.\,Gorbachev$^{a}$, A.\,A.\,Grebeniuk$^{a}$,
D.\,N.\,Grigoriev$^{a,b}$, S.\,K.\,Dhawan$^{d}$,
D.\,A.\,Epifanov$^{a}$,  A.\,S.\,Zaitsev$^{a,b}$,
S.\,G.\,Zverev$^{a}$,  F.\,V.\,Ignatov$^{a}$,
V.\,F.\,Kazanin$^{a,b}$, S.\,V.\,Karpov$^{a}$,
I.\,A.\,Koop$^{a,b}$, P.\,P.\,Krokovny$^{a,b}$,
A.\,S.\,Kuzmin$^{a,b}$, I.\,B.\,Logashenko$^{a,c}$,
P.\,A.\,Lukin$^{a}$, A.\,P.\,Lysenko$^{a}$,
A.\,I.\,Milstein$^{a,b}$, K.\,Yu.\,Mikhailov$^{a}$,
I.\,N.\,Nesterenko$^{a,b}$, M.\,A.\,Nikulin$^{a,b}$,
A.\,V.\,Otboev$^{a}$, V.\,S.\,Okhapkin$^{a}$,
E.\,A.\,Perevedentsev$^{a,b}$, A.\,A.\,Polunin$^{a}$,
A.\,S.\,Popov$^{a}$, S.\,I.\,Redin$^{a}$,
B.\,L.\,Roberts, N.\,I.\,Root$^{a}$,
A.\,A.\,Ruban$^{a}$, N.\,M.\,Ryskulov$^{a}$,
A.\,L.\,Sibidanov$^{a}$, V.\,A.\,Sidorov$^{a}$,
A.\,N.\,Skrinsky$^{a}$, V.\,P.\,Smakhtin,
I.\,G.\,Snopkov$^{a}$, E.\,P.\,Solodov$^{a,b}$,
J.\,A.\,Thompson$^{e\dag}$, G.\,V.\,Fedotovich$^{a,b}$,
B.\,I.\,Khazin$^{a,b}$, V.\,W.\,Hughes$^{d\dag}$,
A.\,G.\,Shamov$^{a}$, Yu.\,M.\,Shatunov$^{a}$,
B.\,A.\,Shwartz$^{a,b}$, S.\,I.\,Eidelman$^{a,b}$,
Yu.\,V.\,Yudin$^{a}$
}
\rauthor{CMD-2 Collaboration}
\address{
{\it $^a$
Budker Institute of Nuclear Physics, Russian Acad. Sci., Siberian Div.,
Novosibirsk, 630090, Russia}\\
{\it $^b$Novosibirsk State University,
Novosibirsk, 630090, Russia}\\
{\it $^c$Boston University, Boston, MA 02215, USA}\\
{\it $^d$Yale University, New Haven, CT 06511, USA}\\
{\it $^e$University of Pittsburgh, Pittsburgh, PA 15260, USA}\\
{\it $^f$Weizmann Institute of Science,  76100, Rehovot, Israel}\\
\vskip 3mm
\rm Published in JETP LETTERS Vol. 82 No. 12 (2005) 743-747
\vskip -10mm
}
\abstract{ The cross section for the process
$e^+e^-\rightarrow\pi^+\pi^-$ is measured in the c.m. energy range
$1.04\div1.38$\,GeV from 995 000 selected collinear events
including 860000 $e^+e^-$ events, 82000 $\mu^+\mu^-$ events, and 33000
$\pi^+\pi^-$ events. The systematic and statistical errors of
measuring the pion form factor are equal to 1.2 $\div$ 4.2 and 5
$\div$ 13\,\%, respectively. } 

\begin{document}
\maketitle
\begin{center}{INTRODUCTION}\end{center}

A study of the cross section for the process
$e^+e^-\rightarrow\pi^+\pi^-$ provides important information on the
electromagnetic form factor of the pion, which describes its internal
structure. Moreover, precision measurement of this cross section is
necessary for calculating the anomalous magnetic moment of the muon
$(g-2)_\mu$~\cite{Davier:2004gb} and its comparison with precision
measurements, one of which was carried out recently at
BNL~\cite{g2}. Such comparison is an important test of the Standard
Model.

\medskip\begin{center}{EXPERIMENT}\end{center}

Measurements were performed at the VEPP-2M collider~\cite{VEPP} with
the CMD-2 general purpose detector (cryogenic magnetic detector), which
combines the properties of a magnetic spectrometer and good
calorimetry~\cite{CMD285,CMD288}. The coordinates, emission angles,
and momenta of charged particles are measured by the coordinate system
of the detector, which consists of the drift and Z chambers located
inside a thin ($0.38X_0$) superconducting solenoid with a magnetic
field of~1\,T. Cylindrical and endcap electromagnetic calorimeters
based on CsI and BGO scintillation crystals ensure measurement of
energy and photon emission angles and make it possible to separate
electrons and hadrons. A range system is used to identify muons.

This work continues a cycle of precision measurements of hadron cross
sections with the CMD-2 detector. The results of the pion form factor
measurement in the energy range 0.61 $\div$ 0.96 GeV were published
in~\cite{Akhmetshin:2001ig}. In this work, we present the results of
the measurement of the form factor in the energy range 1.04 $\div$
1.38~GeV. A more detailed description of the data analysis was given
in~\cite{fpihigh}.

\par An integrated luminosity of 6 pb$^{-1}$ was collected in the
experiment. For analysis, we selected 33 000 $\pi^+\pi^-$ events
accumulated at 35 beam energy points from 520 to 690\,MeV with a step
of 5\,MeV. The beam energy was controlled with an accuracy not worse
than $\delta E/E \sim 10^{-3}$ using the magnetic field value in the
VEPP-2M storage ring.

\medskip\begin{center}{SELECTION OF COLLINEAR EVENTS}\end{center} 
To separate the events $e^+e^-\to e^+e^-$, $e^+e^-\to \mu^+ \mu^-$,
and $e^+e^-\to \pi^+ \pi^-$, the following conditions were used:
\begin{list}{$\bullet$}
{\itemsep 0 mm
\parsep 0 mm
\itemindent 0 mm
\leftmargin \parindent
}
\item One vertex with two tracks of particles with opposite charges
  was found in the drift chamber.
\item The event vertex was located near the beam interaction point;
  i.e., $\rho_{vtx}=$ min$( \rho_{tr}^+, \rho_{tr}^- ) < 0.15$\,cm,
  where $\rho_{tr}^{\pm}$ is the minimum distance between the particle
  track and beam axis, and $|Z_{vtx}|< 10$\,cm, where $Z_{vtx}$ is the
  position of the vertex along the beam axis.
\item Track collinearity conditions:
\item[] $|\Delta \varphi|=|\pi-|\varphi^+-\varphi^-|| < 0.15$, where
  $\varphi^{\pm}$ is the azimuth track angle; 
\item[] $|\Delta \theta|=|\pi-(\theta^++\theta^-)| < 0.25$, where $\theta^{\pm}$ is the polar track angle.
\item Constraint on the solid angle of the event detection: 
\item[] $\theta_{min} < (\pi+\theta^--\theta^+)/2 < \pi -
 \theta_{min}$,  where $\theta_{min}$ = 1.1.
\item The mean momentum was bounded from above to reduce the cosmic
  particle background and from below to suppress $e^+e^-\to K^+K^-$
  events: 
\item[] $E_{beam}+150$\,MeV/c $ > (p^++p^-)/2 > \\ >
  $max$(\sqrt{E_{beam}^2-494^2}\cdot 1.15$\,MeV/c$,300$\,MeV/c$)$,
  where $p^{+}$ and $p^{-}$ are the momenta of the positive and
  negative particles, respectively.
\end{list}

The main sources of the background for the process
$e^+e^-\to\pi^+\pi^-$ are the $e^+e^- \rightarrow
\pi^+\pi^-\pi^0\pi^0$, $e^+e^- \rightarrow \pi^+\pi^-\pi^0$, $e^+e^-
\to K^+K^-$ reactions and cosmic particles. The physical background
contribution to the pion form factor was calculated using experimental
cross sections~\cite{popov,nroot} by taking into account the detection
efficiency determined from the total simulation. The total
contribution of these processes does not exceed 0.8\,\% and is taken
into account as a correction to the pion form factor according to
Eq.\,(\ref{formdef}). The number of cosmic particle background events
was determined from the distribution of the event vertices over the
distance from the beam interaction point.

\begin{figure}[t]
\centering \includegraphics[width=0.48\textwidth]{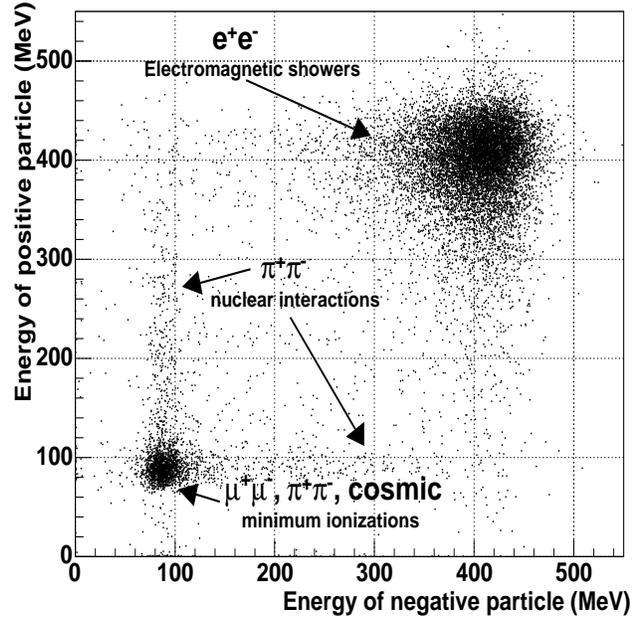}
\caption{Fig. 1. Distribution of collinear particles over the energy deposition in the calorimeter.}
\end{figure}

\medskip\begin{center}{EVENT SEPARATION} \end{center} 
To determine the number of events of each process, we used
two-dimensional distributions over the energy deposition in the CsI
calorimeter (Fig.\,1). 
The number of events ($N_\pi$, $N_\mu+N_e$ ) was determined by
minimizing the maximum likelihood function:
$$\mathcal{L} = - \sum_{events}\ln\Bigl(\sum_{i} N_{i}\cdot
f_{i}(E^+,E^-)\Bigr) + \sum_{i} N_{i},$$ 
where $f_{i}$ is the probability density function for events of a
given type ($\pi,\mu,e,cosmic$). Electrons and positrons initiate an
electromagnetic shower in a calorimeter and thereby noticeably differ
from other charged particles in their energy deposition. The energy
depositions of muons, cosmic particles, and pions with only ionization
losses are almost identical. For this reason, to determine the number
of muons, we used additional information on the ratio of the cross
section for muon production to the cross section for $e^+e^-\to
e^+e^-$ as obtained from the QED calculation with the inclusion of the
resolutions and detection efficiencies in the detector.

\medskip\begin{center}{DETERMINATION OF THE PION FORM FACTOR}\end{center}
The cross section for the $e^+e^-\rightarrow\pi^+\pi^-$ process
integrated over the detection solid angle is given by the expression:
\begin{equation}
\begin{split}
&\sigma_{\pi^+\pi^-} = 
\sigma^{0}_{\pi^+\pi^-}\cdot|F_\pi|^2 \\
&=
\frac{\pi\alpha^2}{3s}\Bigl(1-\frac{4m^2_\pi}{s}\Bigr)^\frac{3}{2}
\frac{3\cos\theta_{min}-\cos^3\theta_{min}}{2}
\cdot|F_\pi|^2,\\
\end{split}
\end{equation}
where $\sigma^{0}_{\pi^+\pi^-}$ is the cross section calculated under
the assumption of the absence of the internal structure of the
pion. The contribution of vacuum polarization to the photon propagator
is also included in the form factor. 

\par At each energy point, the form factor is calculated as:
\begin{multline}
\label{formdef}
\displaystyle
\left| F_{\pi }\right| ^{2}=
{\bf \frac{N_{\pi^+ \pi^- }}{N_{e^+e^-}+N_{\mu^+ \mu^- }}}
\times \\
\times\frac{
\sigma ^{0}_{e^+e^-}\cdot (1+\delta_{e^+e^-})\cdot\varepsilon_{e^+e^-}
+\sigma ^{0}_{\mu^+\mu^-}\cdot (1+\delta_{\mu^+\mu^-})\cdot\varepsilon_{\mu^+\mu^-}}
{\sigma ^{0}_{\pi^+\pi^-}\cdot (1+\delta_{\pi^+\pi^-})
\cdot\varepsilon_{\pi^+\pi^-}\cdot(1-\Delta_{\pi\; loss} )} \\
-\Delta _{3\pi,4\pi,K^+K^-},
\end{multline}
Here, $N_{\pi^+ \pi^- }/(N_{e^+e^-}+N_{\mu^+ \mu^- })$ is the ratio of
the number of detected pions to the number of muons and electrons as
obtained in the minimization procedure, $\sigma^{0}_{ii}$ is the Born
cross section in the lowest order of perturbation theory,
$\delta_{ii}$ is the radiative correction, $\varepsilon_{ii}$ is the
detection efficiency, $\Delta_{3\pi,4\pi,K^+K^-}$ is the correction
for background processes, and $\Delta_{loss}$ is the correction for
the pion loss at the vacuum chamber wall and drift chamber material
due to nuclear interactions. The $\Delta_{loss}$ correction was
determined from simulation by comparing the number of selected pions
with the inclusion and exclusion of nuclear interactions. The
correction value was equal to $0.8\div1.2\,\%$.

Collinear events were selected using only information from the drift
chamber. For this reason, by selecting desired (test) events from the
CsI calorimeter and checking whether the reconstructed tracks are in
the drift chamber, one can determine the event reconstruction
efficiency. The reconstruction efficiency was
$97\div98\,\%$. According to Eq.\,(\ref{formdef}), only the
difference between the detection efficiencies for different processes,
which was $0.16\pm0.09\,\%$ between electrons and muons, is important
for the determination of the pion form factor.

The radiative corrections for the processes $e^+e^-\rightarrow
e^+e^-$, $e^+e^-\rightarrow \mu^+\mu^-$, and $e^+e^-\rightarrow
\pi^+\pi^-$ are calculated using a procedure that was developed
in~\cite{radcor} and was based on the formulas
from~\cite{rcpp,rcee}. The probability of the emission of many photons
along the initial and final particles, the emission of one photon at a
large angle, and the vacuum polarization contribution to the photon
propagator are taken into account in these formulas. Since the vacuum
polarization is usually included in the definition of the form factor,
its contribution was not included in the radiative correction for the
process $e^+e^-\rightarrow \pi^+\pi^-$. According to~\cite{radcor},
the accuracy of the calculation of the radiative corrections is
estimated as 0.2\,\% for all the processes. The number of selected
collinear events depends on the angular and momentum resolutions of
the drift chamber. In order to include them in the calculation for
radiative corrections, the emission angles and momenta of particles
were additionally simulated according to the experimental resolution
and then the selection criteria were imposed. For the
$e^+e^-\rightarrow e^+e^-$ process, the bremsstrahlung energy losses
of electrons and positrons on the vacuum chamber wall and first 10\,cm
of the drift chamber were taken into account.

Table 1 presents the form factor at each energy point. 

\begin{table}[t]
\caption{\label{table:fpi}Table 1. Experimental pion form factor
  $|F_{\pi}|^2$. Only the statistical error is given}
\begin{center}
\begin{tabular}{|c|c||c|c|}
\hline
 $E,$&  ${|F_{\pi}|}^{2}$ & $E,$ &  ${|F_{\pi}|}^{2}$ \\
MeV & & MeV &    \\
\hline
490.0 & 3.596 $\pm$ 0.163 & 605.7 & 1.069 $\pm$ 0.082 \\ \hline
520.0 & 2.598 $\pm$ 0.134 & 610.0 & 0.989 $\pm$ 0.075 \\ \hline
525.0 & 2.262 $\pm$ 0.112 & 615.0 & 1.069 $\pm$ 0.088 \\ \hline
530.0 & 2.185 $\pm$ 0.135 & 620.0 & 0.988 $\pm$ 0.081 \\ \hline
535.0 & 2.295 $\pm$ 0.130 & 625.0 & 0.794 $\pm$ 0.064 \\ \hline
540.0 & 1.884 $\pm$ 0.119 & 630.0 & 0.696 $\pm$ 0.063 \\ \hline
545.0 & 2.120 $\pm$ 0.110 & 635.0 & 0.719 $\pm$ 0.057 \\ \hline
550.0 & 1.704 $\pm$ 0.120 & 640.0 & 0.693 $\pm$ 0.052 \\ \hline
555.0 & 1.641 $\pm$ 0.106 & 645.0 & 0.571 $\pm$ 0.042 \\ \hline
560.0 & 1.449 $\pm$ 0.146 & 650.0 & 0.640 $\pm$ 0.046 \\ \hline
565.0 & 1.683 $\pm$ 0.103 & 655.0 & 0.570 $\pm$ 0.050 \\ \hline
570.0 & 1.531 $\pm$ 0.088 & 660.0 & 0.483 $\pm$ 0.054 \\ \hline
575.0 & 1.374 $\pm$ 0.150 & 665.0 & 0.460 $\pm$ 0.040 \\ \hline
580.0 & 1.386 $\pm$ 0.087 & 670.0 & 0.524 $\pm$ 0.062 \\ \hline
585.0 & 1.197 $\pm$ 0.115 & 675.0 & 0.347 $\pm$ 0.049 \\ \hline
590.0 & 1.200 $\pm$ 0.088 & 680.0 & 0.357 $\pm$ 0.040 \\ \hline
595.0 & 1.014 $\pm$ 0.093 & 685.0 & 0.424 $\pm$ 0.078 \\ \hline
600.0 & 0.983 $\pm$ 0.079 & 690.0 & 0.338 $\pm$ 0.032 \\ \hline
\end{tabular}
\end{center}
\end{table}

\medskip\begin{center}{SYSTEMATIC ERROR}\end{center} 
The main contributions to the systematic error are listed in
Table~2. The systematic error increases with energy, because the error
in the number of muons directly contributes to the error in the
number of pions, and the ratio of the number of muons to the number of
pions increases from 1 to 7 when the c.m. energy increases from 1 to
1.38 GeV. The total systematic error is equal to $1.2\div4.2\,\%$ and
does not exceed one third of the statistical error at each experimental
point.

One of the tests of the separation procedure was performed using the
simulation of $e^+e^-$, $\mu^+\mu^-$, and $\pi^+\pi^-$ events. The
simulation data were analyzed with the inclusion of the following
corrections: pion loss due to the nuclear interaction, the energy
losses of electrons at the vacuum-chamber wall, and the resolution of
the drift chamber while calculating radiative corrections. The
calculated difference between the detection efficiencies for $e^+e^-$
and $\mu^+\mu^-$,$\pi^+\pi^-$ in the simulation was equal to
$\varepsilon_{MIP}-\varepsilon_{e^+e^-}=0.189\pm0.004\,\%$ in good
agreement with the measured value. The difference between the form
factor obtained and the form factor used in the simulation varies from
0.2 to 1.5\,\% in dependence on the energy. The difference at the
highest energy was equal to 1.5\,\% and consisted of 1\,\% for the
separation procedure and 0.5\,\% characterizing the systematic error
in the inclusion of the above corrections.

\begin{table}[t]
\centering
\caption{\label{systfpi}Table 2. Various contributions to the
 systematic error in $|F_{\pi}|^2$. The given range corresponds to the scanned
 energy range}
\vskip 0.1 in
\begin{tabular}{l|c} \hline
  Error source & Error \\
%\multicolumn{2}{r}{$\sqrt{s}$= 1.04 $\div$ 1.38 GeV}\\ \hline
&$\sqrt{s}$= 1.04 $\div$ 1.38 GeV\\ \hline
Detection solid angle         & 0.2$\div$0.5 \% \\	 
Detection efficiency 	      & 0.5$\div$ 2 \% \\	 
Pion loss 		      & 0.2 \% \\		 
Bremsstrahlung $e^+e^-$      & 0.05$\div$1.7 \% \\	 
Radiative corrections 	      & 0.5$\div$ 2 \% \\	 
Background events 	      & 0.6$\div$1.6 \% \\	 
Energy calibration 	      & 0.7$\div$1.1 \% \\	 
Particle separation procedure & 0.2$\div$1.5 \%  \\\hline
			      & 1.2$\div$4.2 \% \\	 
Statistical error at the point& 5$\div$13 \% \\  \hline         
\end{tabular}
\end{table}

\medskip\begin{center}{DISCUSSION}\end{center}
Figure 2 shows the results, which are in good agreement with the data
obtained in the previous experiments with the detectors
OLYA\cite{cmd_olya}, DM1\cite{Quenzer:1978qt},
DM2\cite{Bisello:1988hq}, BCF\cite{Bollini:1975pn},
ACO\cite{Cosme:1976ft}. The form factor in this energy range was
measured in detail only in the experiment with the OLYA detector with
a systematic error of $10\div15$\,\%. The experimental energy
dependence of the form factor is well reproduced in the framework of
the vector-meson dominance model by the sum of the amplitudes of the
$\rho(770)$, $\rho(1450)$, $\rho(1700)$, $\omega$ and $\phi$
mesons~\cite{Akhmetshin:2001ig}:
\begin{multline}
\label{func:GS}
|F_\pi(s)|^2=\\
\biggr|\Bigr(\mathrm{BW}^{\mathrm{GS}}_{\rho(770)}(s)\cdot
\displaystyle \bigr(\mathstrut 1+\delta_{\omega}  \frac{s}{m_\omega^2}\,
\mathrm{BW}_{\omega}(s)+\delta_{\phi}  \frac{s}{m_\phi^2}\, \mathrm{BW}_{\phi}(s)\bigl)+ \\
+ \beta \, \mathrm{BW}^{\mathrm{GS}}_{\rho(1450)}(s) + \gamma \, \mathrm{BW}^{\mathrm{GS}}_{\rho(1700)}(s)
\Bigl)/(1+\beta+\gamma)\biggl|^2,
\end{multline}
Here, $\mathrm{BW}^{\mathrm{GS}}_{\rho}(s)$ is the meson
parametrization in the Gounaris-Sakurai model~\cite{gunsac};
$\mathrm{BW}_{\omega}(s)$ and $\mathrm{BW}_{\phi}(s)$ are the
parametrization of the $\omega$ and $\phi$ resonances, respectively,
which were represented by the relativistic Breit-Wigner form due to
a small width; $\delta_{\omega}$,$\delta_{\phi}$, $\beta$ and $\gamma$
are the model parameters describing the relative contributions of the
$\rho-\omega$ and $\rho-\phi$ interferences and $\rho(1450)$ and
$\rho(1700)$ states, respectively. In order to determine the model
parameters, it is necessary to use all the available data on the form
factor in the energy range $\sqrt{s}=0.36\div3.7$\,GeV, which will be
done in a future work with analysis of all the information accumulated
at the CMD-2 detector in the energy range from 0.37 to 1.38\,GeV.

\begin{figure}[t]
\centering \includegraphics[width=0.48\textwidth]{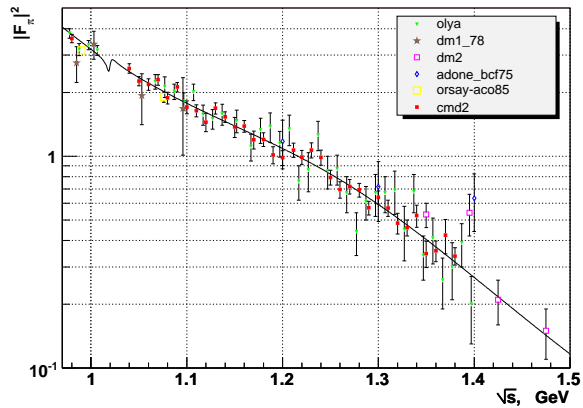}
\caption{ Fig. 2. Experimental data obtained for the pion form factor
$|F_{\pi}|^2$ in this work in comparison with other experiments.}
\end{figure}

\medskip\begin{center}{CONCLUSIONS} \end{center}
In this work, the cross section for the process
$e^+e^-\rightarrow\pi^+\pi^-$ was measured in the c.m. energy range
$1.04\div1.38$\,GeV with the world best accuracy. The
systematic and statistical errors of the measurement are equal to
$1.2\div 4.2\,\%$ and $5\div13\,\%$, respectively. The measured cross
section agrees well with the results of the previous experiments.

This work was supported by the Russian Foundation for Basic Research
(project nos. 03-02-16477, 03-0216280, 03-02-16843, 04-02-16217,
04-02-16223, and 04-02-16434).

\end{document}